\documentclass[12pt, a4paper]{article}

\usepackage{amsmath,amssymb,amsthm}
\usepackage{graphicx}
\usepackage[left=3.cm, top=4.0cm, right=3.cm, bottom=4.0cm]{geometry}
\usepackage[english]{babel}

\usepackage{hyperref}

\newcommand{\ud}{\mathrm{d}}
\newcommand{\Tr}{\,\mathrm{Tr}}
\newcommand{\I}{\mathrm{i}}
\newcommand{\N}{\mathbb{N}}

\newcommand{\RNum}[1]{\uppercase\expandafter{\romannumeral #1\relax}}
\allowdisplaybreaks[4]

\newcommand{\Hlog}{\mathrm{Hlog}}



\newtheoremstyle{mydefinition}{3pt}{3pt}{\sf}{}{\bf}{.}{.5em}{}

\theoremstyle{mydefinition}
\newtheorem{dfnt}{Definition}%[section]%[section]/[chapter]

\theoremstyle{plain}
\newtheorem{lemma}[dfnt]{Lemma}%[section]%[section]/[chapter]
\newtheorem{thrm}{Theorem}%[section]%[section]/[chapter]
\newtheorem{cor}[dfnt]{Corollary}

\theoremstyle{remark}
\numberwithin{dfnt}{section}

\begin{document}

\thispagestyle{plain}
\begin{center}

{\Large
\textbf{Solution of the self-dual $\Phi^4$ QFT-model on \\[0.5ex]
four-dimensional Moyal space}*\footnotetext[1]{With an 
appendix by Robert Seiringer, Institute of Science and Technology Austria,
Am Campus 1, 3400 Klosterneuburg, Austria}
} 

\par
\vspace*{4ex}

{\large Harald Grosse$^{1,3}$, Alexander Hock$^{2}$, Raimar
  Wulkenhaar$^2$}

%\vspace*{1ex}

%{\large With an appendix by Robert Seiringer$^4$}

\vspace*{2ex}

\emph{\small$^1$ Fakult\"at f\"ur Physik, Universit\"at Wien,
        Boltzmanngasse 5, A-1090 Vienna, Austria}
\\[0.5ex]
\emph{\small$^2$ Mathematisches Institut der
        Westf\"alischen Wilhelms-Universit\"at,\\
Einsteinstra\ss{}e 62, D-48149 M\"unster, Germany}
\\[0.5ex]
\emph{\small$^3$ Erwin Schr\"odinger International Institute
for Mathematics and Physics, \\ University of Vienna,
        Boltzmanngasse 9, A-1090 Vienna, Austria}
%\\[0.5ex]
%\emph{\small$^4$ Institute of Science and Technology Austria,
%Am Campus 1, 3400 Klosterneuburg, Austria}
\end{center}

{\raggedright
\small{E-mails: harald.grosse@univie.ac.at, a\_hock03@wwu.de,
  raimar@math.uni-muenster.de
  %, robert.seiringer@ist.ac.at
  }
\vspace*{4ex}
\hrule \vspace*{2ex}}

\noindent
\textbf{Abstract}
\\
Previously the exact solution of the planar sector of the self-dual
$\Phi^4$-model on 4-dimensional Moyal space was established up to the
solution of a Fredholm integral equation. This paper solves, for any
coupling constant $\lambda>-\frac{1}{\pi}$, the Fredholm equation in
terms of a hypergeometric function and thus completes the construction of
the planar sector of the model. We prove that the interacting model
has spectral dimension $4-2\frac{\arcsin(\lambda\pi)}{\pi}$ for
$|\lambda|<\frac{1}{\pi}$. It is this dimension drop which for
$\lambda>0$ avoids the triviality problem of the matricial
$\Phi^4_4$-model.

We also establish the power series approximation of the Fredholm
solution to all orders in $\lambda$.
The appearing functions are hyperlogarithms defined by iterated integrals,
here of alternating letters $0$ and $-1$. We identify the
renormalisation parameter
which gives the same normalisation as the ribbon graph expansion.

\vspace*{2ex}
\hrule
\vspace*{2ex}

\\
\begin{tabular}{@{}p{0.14\textwidth}@{}p{0.86\textwidth}@{}}
Keywords: & Solvable models, noncommutative QFT, special functions
\end{tabular}

\normalsize

\section{Introduction}

Many quantum field theory models have been solved or constructed
in two dimensions, see e.g.\
\cite{Thirring:1958in,Schwinger:1962tp,Gross:1974jv}.  Until now there
is nothing comparable in four dimensions. The perturbative
renormalisation of the $\Phi^4$-model on four-dimensional Moyal space with
harmonic propagation \cite{Grosse:2004yu} and the proof that
the $\beta$-function vanishes \cite{Disertori:2006nq} at a self-duality
point provided some hope to construct this particular
four-dimensional model.

At a special self-duality point \cite{Langmann:2002cc}, the model reduces
to a dynamical matrix model with action
\begin{align}
 S[\Phi]=V\Tr(E\Phi^2)+\frac{\lambda}{4} V\Tr(\Phi^4) \label{action}
\end{align}
for self-adjoint $\mathcal{N}{\times}\mathcal{N}$-matrices $\Phi$,
where $E$ has eigenvalues
$E_n=\frac{n}{\sqrt{V}}+\frac{\mu^2_{bare}}{2}$ which arise with multiplicity
$n$. The parameter $V\in \mathbb{R}$ is the deformation parameter of the
Moyal space, $\lambda\in\mathbb{R}$ is the coupling constant and
$\mu^2_{bare}$ the unrenormalised mass square. The action $S[\Phi]$
is employed to define correlation functions
\begin{align*}
 \langle \Phi_{a_1b_1}\Phi_{a_2b_2} \dots \Phi_{a_nb_n}\rangle :=
\log  \Big(\frac{\int \ud \Phi\, \Phi_{a_1b_1}\Phi_{a_2b_2} \cdots
\Phi_{a_nb_n} e^{-S[\Phi]} }
 {\int \ud \Phi\,  e^{-S[\Phi]}}\Big).
\end{align*}
Integration by parts produces many relations between these correlation
functions. Further relations result from a Ward-Takahashi identity
discovered in \cite{Disertori:2006nq}. It was shown in
\cite{Grosse:2012uv} that these relations can be organised into a
closed non-linear equation for the planar two-point function and a
hierarchy of Dyson-Schwinger equations for all other functions.
The latter are linear in the function of interest with an inhomogeneity
that only depends on finitely many functions known by induction.

As characteristic to matrix models, the two-point function has
a formal genus expansion
\[
\langle \Phi_{ab}\Phi_{ba}\rangle=\sum_{g=0}^\infty
V^{1-2g} Z G_{ab}^{(g)}\;.
\]
Its planar part $G_{ab}^{(0)}$ can be isolated in a
limit $V\to \infty$. Particularly transparent is a combined limit
where also the size $\mathcal{N}$ of the matrices is sent to $\infty$, with the
ratio $\frac{\mathcal{N}}{\sqrt{V}}=\Lambda^2$ fixed.
The previously discrete eigenvalues $E_n$ become in this limit functions
$E_x=x+\frac{\mu_{bare}^2}{2}$ of a real variable $x\in [0,\Lambda^2]$, and
$G_{ab}^{(0)}$ converges to $G(x,y)$ with
$x=\lim \frac{a}{\sqrt{V}}$ and $y=\lim \frac{b}{\sqrt{V}}$.
It this setting the Dyson-Schwinger equation for $G_{ab}^{(0)}$ converges to
a non-linear integral equation \cite{Grosse:2012uv}
\begin{align}
\label{inteq}
\Big(\mu_{bare}^2+x+y
+ \lambda \int_0^{\Lambda^2} \!\! dt \;
t\,ZG(x,t)\Big)Z G(x,y)
&= 1+\lambda
\int_0^{\Lambda^2} \!\! dt \; t\,Z\frac{G(t,y)-G(x,y)}{t-x}.
\end{align}
It is understood that $\mu_{bare}^2$ and $Z$ depend on the cut-off $\Lambda$.
According to the renormalisation philosophy, the task is to determine the
precise dependence $\mu_{bare}^2(\Lambda)$, $Z(\Lambda)$ so that the solution
$G(x,y)$ of (\ref{inteq}) has a limit $\Lambda\to \infty$.

In our recent work \cite{Grosse:2019jnv} we succeeded in solving
the analogue of (\ref{inteq}) for general eigenvalues $E_a$ and without
requiring the special limit $\mathcal{N},V\to \infty$, up to
the determination of an implicitly defined measure function.
In case of (\ref{inteq}) this solution specifies to:
\begin{thrm}[\cite{Grosse:2019jnv}]\label{Thrm1}
Equation \eqref{inteq} for the renormalised planar 2-point function
of the $\phi^4$ QFT-model on four-dimensional noncommutative Moyal space
is solved by
 \begin{align*}
 G(x,y)&=\frac{\mu^2 \exp(N(x,y))}{\mu^2+x+y},
\\
 N(x,y)&:=\frac{1}{2\pi \I}\int_{-\infty}^\infty \!\! dt\:
\bigg\{\log\big(x-J(-\tfrac{\mu^2}{2}-\I t)\big)\frac{d}{dt}
 \log\big(y-J(-\tfrac{\mu^2}{2}+\I t)\big)
\\
 & \qquad\qquad\qquad
-\log\big(-J(-\tfrac{\mu^2}{2}-\I t)\big)\frac{d}{dt}
 \log\big(-J(-\tfrac{\mu^2}{2}+\I t)\big)
\\
 &\qquad\qquad\qquad
-\log\big(x-(-\tfrac{\mu^2}{2}-\I t)\big)\frac{d}{dt}
 \log\big(y-(-\tfrac{\mu^2}{2}+\I t)\big)
\\
&\qquad\qquad\qquad
+\log\big(-(-\tfrac{\mu^2}{2}-\I t)\big)\frac{d}{dt}
 \log\big(-(-\tfrac{\mu^2}{2}+\I t)\big)\bigg\},
\end{align*}
where $J$ is the solution of a Fredholm integral equation of second kind:
\begin{align}\label{Fred}
 J(x)=x-\lambda x^2\int_0^\infty \!\! dt\;\frac{J(t)}{(t+\mu^2)^2(t+\mu^2+x)}\;.
\end{align}
Here $\mu>0$ is a free renormalisation parameter, and $G(0,0)=1$ is already implemented.
\end{thrm}
As main result of this paper we prove that (\ref{Fred}) is solved by
a hypergeometric function,
\begin{align}
\label{Jx-final}
 J(x)=x  \;_2F_1\Big(\genfrac{}{}{0pt}{}{
\alpha_\lambda,\;1-\alpha_\lambda}{2}\Big|-\frac{x}{\mu^2}\Big),\qquad
\text{where} \quad
\alpha_\lambda:=\frac{\arcsin(\lambda\pi)}{\pi}.
\end{align}
Moreover, we show that the particular choice
$\mu^2=\frac{\alpha_\lambda(1-\alpha_\lambda)}{\lambda}$ provides the
same normalisation as the expansion into renormalised ribbon graphs.

The following sections present several methods which we employed to
find the solution \eqref{Jx-final} of \eqref{Fred}.  In
sec.~\ref{sec:DGL} we show that a rescaling of $J$ satisfies a
hypergeometric differential equation from which we deduce
\eqref{Jx-final}. Some steps rely on Appendix \ref{app} where the spectrum
of an integral operator is determined.  
In subsection \ref{sec:spec} we determine the spectral dimension.
The treatment via a differential equations is probably the most
elegant one. We first obtained this solution
via a perturbative expansion described in sec.~\ref{Sec:perturbative}.
We understand the pattern of the power series solution of \eqref{Fred}
to $\mathcal{O}(\lambda^{10})$ and resum it to
\eqref{Jx-final}. The advantage of this approach is that it identifies the
renormalisation parameter $\mu^2$ for which our solution matches the
usual perturbative renormalisation prescription.
Finally, in sec.~\ref{Sec:proof} we directly verify
\eqref{Jx-final} via integrals for Meijer-G functions.

\section{Solution via differential equation}

\label{sec:DGL}

It is convenient to symmetrise the Fredholm equation \eqref{Fred}.
Dividing by $\frac{x}{\mu^2+x}$ and defining
$\tilde{\varrho}_\lambda(x):=\frac{J(x)}{x(\mu^2+x)}$, we have
\begin{align}
 \tilde{\varrho}_\lambda(x)&=\frac{1}{\mu^2+x}
-\lambda \int_0^\infty \!\! dt\;\frac{\tilde{\varrho}_\lambda(t)\,tx}{(\mu^2+t)
 (\mu^2+x)(\mu^2+x+t)}
\nonumber
\\*
&= \frac{c_\lambda}{\mu^2+x}-\lambda \int_0^\infty \!\! dt\;
 \frac{\tilde{\varrho}_\lambda(t)}{\mu^2+x+t},\label{intc}
\end{align}
where $c_\lambda =1+\lambda\mu^2\int_0^\infty dt\;
\frac{\tilde{\varrho}_\lambda(t)}{\mu^2+t}=
1+\lambda\mu^2\int_0^\infty dt\;\frac{J(t)}{t(\mu^2+t)^2}$.
The second line results by (not so obvious) rational fraction expansion.
As proved in appendix~\ref{app},
there exists for $\lambda>-\frac{1}{\pi}$
a solution $\tilde{\varrho}_\lambda\in L^2(\mathbb{R}_+)$,
which means $\lim_{t\to \infty} t \tilde{\varrho}_\lambda (t)=0$.
Another transformation $\phi(x)=\mu^2\tilde{\varrho}_\lambda(x\mu^2)$
simplifies the problem to
\begin{align}\label{feq}
\phi(x)=\frac{c_\lambda}{1+x}
-\lambda \int_0^\infty \!\! dt\; \frac{\phi(t)}{1+t+x},\qquad
\phi(0)=1.
\end{align}
The aim is to find the differential operator $D_x$ acting on
\eqref{feq} which is reproduced under the integral on $\phi(t)$ such
that all appearing inhomogeneous parts vanish, i.e.
\begin{align*}
 D_x\phi(x)=-\lambda\int_0^\infty dt\frac{D_t \phi(t)}{1+t+x}.
\end{align*}
We compute derivatives and integrate by parts, taking the boundary 
values at $0$ and $\infty$ into account:
\begin{align}
\phi'(x)=-\frac{c_\lambda}{(1+x)^2}
+\lambda\int_0^\infty \!\! dt\;\frac{\phi'(t)}{1+t+x}+\frac{\lambda}{1+x}.
\label{dphi}
\end{align}
Also the product with $1+x$ simplifies by integration by parts:
\begin{align}
(1+x)\phi'(x)&=-\frac{c_\lambda}{(1+x)}
-\lambda\int_0^\infty \!\! dt\;\frac{t \phi'(t)}{1+t+x}
\;.
\label{xdphi}
\end{align}
We differentiate once more:
\begin{align}
(1+x)\phi''(x)
+ \phi'(x)
&=\frac{c_\lambda}{(1+x)^2}
+\lambda\int_0^\infty \frac{dt}{(1+t+x)}\,
\frac{d}{dt}( t\phi'(t))\;,
\nonumber
\\
(1+x)\phi''(x)
&=\frac{2c_\lambda}{(1+x)^2}
+\lambda\int_0^\infty \!\! dt\;\frac{t \phi''(t)}{1+t+x}
-\frac{\lambda}{1+x}\;.
\end{align}
We multiply by $x$ and integrate by parts:
\begin{align*}
x(1+x)\phi''(x)
&=\frac{2c_\lambda}{(1+x)}-\frac{2c_\lambda}{(1+x)^2}
-\lambda\int_0^\infty \frac{dt\;  t(1+t) \phi''(t)}{1+t+x}
+\frac{\lambda}{1+x}\;.
\end{align*}
We subtract twice \eqref{dphi} and add four times \eqref{xdphi}:
\begin{align*}
x(1+x)\phi''(x)+(2+4x) \phi'(x)
&=-\frac{2c_\lambda+\lambda}{(1+x)}
-\lambda\int_0^\infty \!\!\! dt\;\frac{t(1+t) \phi''(t)+ (2+4t)\phi'(t)
}{1+t+x}.
\end{align*}
Finally, we add $\frac{2c_\lambda+\lambda}{c_\lambda}$ times
\eqref{feq} to get $D_x=x(1+x)\frac{d^2}{dx^2}+(2+4x)\frac{d}{dx}+\frac{2c_\lambda+\lambda}{c_\lambda}$, or equivalently
\begin{align}
0&=(\mathrm{id}+\lambda \hat{A}_1) g,\qquad \text{where}
\label{idA}
\\
g(x)&=x(1+x)\phi''(x)+(2+ 4x) \phi'(x)+ \frac{2c_\lambda+\lambda}{c_\lambda}
\phi(x),
\nonumber
\end{align}
and $\hat{A}_\mu$ is the integral operator with kernel
$\hat{A}_\mu(t,u)=\frac{1}{u+t+\mu^2}$. The arguments given in
appendix \ref{app} show that $\hat{A}_\mu$ has
spectrum $[0,\pi]$ for any $\mu\geq 0$. Therefore, equation
\eqref{idA} has for
$\lambda>-\frac{1}{\pi}$ only the trivial solution $g(x)=0$, which is
a standard hypergeometric differential equation.
The normalisation $\phi(0)=1$ uniquely fixes the solution to
\begin{align}
\phi(x)&=
{}_2F_1\Big(\genfrac{}{}{0pt}{}{1{+}\alpha_\lambda,\;2{-}\alpha_\lambda}{2}
\Big| -x\Big)
\nonumber
\\
&=\frac{1}{1+x} {}_2F_1\Big(\genfrac{}{}{0pt}{}{\alpha_\lambda,\;1{-}\alpha_\lambda}{2}
\Big| -x\Big)
\;,\quad  c_\lambda=\frac{\lambda}{\alpha_\lambda(1{-}\alpha_\lambda)}.
\label{phi-sol}
\end{align}

It remains to satisfy the boundary condition
$c_\lambda=1+\lambda \int_0^\infty dt\;\frac{\phi(t)}{1+t}$
given after (\ref{intc}).
The integral can be evaluated via the Euler integral
 \cite[\S 9.111]{gradshteyn2007},
\begin{align*}
\int_0^\infty \!\!\! dt\;\frac{\phi(t)}{1+t}
&= \frac{\Gamma(2)}{\Gamma(1-\alpha_\lambda)\Gamma(1+\alpha_\lambda)}
\int_0^\infty dt
\int_0^1 du \;\frac{u^{-\alpha_\lambda}(1-u)^{\alpha_\lambda}}{(1+ut)^{\alpha_\lambda}(1+t)^2}
\nonumber
\\
&= \frac{1}{\Gamma(1-\alpha_\lambda)\Gamma(1+\alpha_\lambda)}
\int_0^1 ds
\int_0^1 du \;\frac{u^{-\alpha_\lambda}(1-u)^{\alpha_\lambda}
(1-s)^{\alpha_\lambda}}{(1-(1-u)s)^{\alpha_\lambda}}
\nonumber
\\
&= \frac{1}{\Gamma(1-\alpha_\lambda)\Gamma(2+\alpha_\lambda)}
\int_0^1 du \;u^{-\alpha_\lambda}(1-u)^{\alpha_\lambda}\;
{}_2F_1\Big(\genfrac{}{}{0pt}{}{\alpha_\lambda,\;1}{2+\alpha_\lambda}
\Big|1-u\Big)
\nonumber
\\
&= \frac{1}{\Gamma(1-\alpha_\lambda)\Gamma(2+\alpha_\lambda)}
\int_0^1 du \;u^{\alpha_\lambda}(1-u)^{-\alpha_\lambda} \;
\Big\{
\frac{(1+\alpha_\lambda)}{\alpha_\lambda}
{}_2F_1\Big(\genfrac{}{}{0pt}{}{\alpha_\lambda,\;1}{1+\alpha_\lambda}
\Big|u\Big)
\nonumber
\\
&\qquad\qquad\qquad -\frac{1}{\alpha_\lambda}
{}_2F_1\Big(\genfrac{}{}{0pt}{}{\alpha_\lambda,\;2}{2+\alpha_\lambda}
\Big|u\Big)
\Big\}
\nonumber
\\
&= \frac{1}{\Gamma(1-\alpha_\lambda)\Gamma(2+\alpha_\lambda)}
\Big\{
\frac{(1+\alpha_\lambda)}{\alpha_\lambda}
\frac{\Gamma(1+\alpha_\lambda)\Gamma(1-\alpha_\lambda) \Gamma(1-\alpha_\lambda)}{
\Gamma(2-\alpha_\lambda)\Gamma(1)}
\nonumber
\\
&\qquad\qquad\qquad -\frac{1}{\alpha_\lambda}
\frac{\Gamma(2+\alpha_\lambda)\Gamma(1+\alpha_\lambda) \Gamma(1-\alpha_\lambda)\Gamma(1-\alpha_\lambda)}{
\Gamma(2)\Gamma(2)\Gamma(1)}
\Big\}
\nonumber
\\
&= \frac{1}{\alpha_\lambda(1-\alpha_\lambda)} -\Gamma(\alpha_\lambda)\Gamma(1-\alpha_\lambda).
\end{align*}
Here we have transformed $t=\frac{s}{1-s}$, evaluated first
the $s$-integral \cite[\S 9.111]{gradshteyn2007}
to a hypergeometric function, used its contiguous relation
\cite[\S 9.137.17]{gradshteyn2007} so that the remaining integrals are
known from \cite[\S 7.512.4]{gradshteyn2007} and
\cite[\S 7.512.3]{gradshteyn2007}.
We thus conclude
\begin{align*}
c_\lambda= 1 +
\frac{\lambda}{\alpha_\lambda(1-\alpha_\lambda)}- \frac{\lambda \pi}{\sin (\alpha_\lambda \pi)}
\stackrel{!}{=}\frac{\lambda}{\alpha_\lambda(1-\alpha_\lambda)}
\end{align*}
with solution
\begin{align}
\sin(\alpha_\lambda\pi)=\lambda\pi\;,\qquad
\alpha_\lambda=\left\{
\begin{array}{cl}
\dfrac{\arcsin(\lambda \pi)}{\pi} & \text{for }
|\lambda| \leq \frac{1}{\pi} ,\\
\dfrac{1}{2}+\mathrm{i} \dfrac{\mathrm{arcosh}(\lambda \pi)}{\pi} & \text{for }
\lambda \geq \frac{1}{\pi}.
\end{array}\right.
\label{sol-alpha}
\end{align}
The branch is uniquely selected by the requirement
$\lim_{\lambda\to 0} c_\lambda =1$. For $\lambda<-\frac{1}{\pi}$ there is
no solution for which $c_\lambda$ and $\phi$ are real.
Transforming back to $\tilde{\rho}_\lambda$ and $J$ gives
the result announced in (\ref{Jx-final}), which provides the
two-point function $G(x,y)$ via
Thm.~\ref{Thrm1}.

\subsection{Spectral dimension}
\label{sec:spec}

Let $\varrho_0(x)dx$ be the spectral measure of the
operator $E$ in the initial action \eqref{action}. The main discovery
of \cite{Grosse:2019jnv} was that the interaction
$\frac{\lambda}{4} \mathrm{Tr}(\Phi^4)$ effectively modifies the
spectral measure to $\varrho_\lambda(x)dx$. What before, when expressed
in terms of $\varrho_0(x)dx$, was intractable became suddenly exactly
solvable in terms of the deformation $\varrho_\lambda(x)dx$.
For four-dimensional Moyal space one has
$\varrho_0(x)=x$ and $\varrho_\lambda(x)=J(x)$. The explicit solution
\eqref{Jx-final} shows that the deformation is
drastic: it changes the spectral dimension $D$
defined by
$D=\inf\{p\;:~ \int_0^\infty dt \;\frac{\varrho_\lambda(t)}{(1+t)^{p/2}}
<\infty\}$.

\begin{lemma}
\label{lem:bound}
For any $|\alpha_\lambda|<\frac{1}{2}$ one has
\[
\frac{1}{(1+x)^{\alpha_\lambda}} \leq
{}_2F_1\Big(\genfrac{}{}{0pt}{}{\alpha_\lambda,\;1{-}\alpha_\lambda}{2}
\Big| -x\Big)
\leq \frac{\Gamma(1-2\alpha_\lambda)}{
\Gamma(2-\alpha_\lambda)\Gamma(1-\alpha_\lambda)}
\frac{1}{(1+x)^{\alpha_\lambda}}\;.
\]
\end{lemma}
\noindent
\emph{Proof.} We transform with \cite[\S 9.131.1]{gradshteyn2007} to
\[
{}_2F_1\Big(\genfrac{}{}{0pt}{}{\alpha_\lambda,\;1{-}\alpha_\lambda}{2}
\Big| -x\Big)
=\Big(\frac{1}{1+x}\Big)^{\alpha_\lambda}\;
\frac{\displaystyle
{}_2F_1\Big(\genfrac{}{}{0pt}{}{2-\alpha_\lambda,\;1{-}\alpha_\lambda}{2}
\Big| \frac{x}{1+x}\Big) }{\displaystyle
\Big(1-\frac{x}{1+x}\Big)^{2\alpha_\lambda-1}}.
\]
By \cite[Thm. 1.10]{ponnusamy_vuorinen_1997}, the fraction on the rhs
is strictly increasing from 1 at $x=0$ to its limit
$\frac{B(2,1-2\alpha_\lambda)}{B(2-\alpha_\lambda,1-\alpha_\lambda)}
=\frac{\Gamma(1-2\alpha_\lambda)}{
\Gamma(2-\alpha_\lambda)\Gamma(1-\alpha_\lambda)}$
for $x\to \infty$. \hspace*{\fill}$\square$%

\begin{cor}
For $|\lambda|<\frac{1}{\pi}$, the deformed measure
$\varrho_\lambda=J$ of four-dimensional Moyal space has spectral dimension
$D=4-2\frac{\arcsin(\lambda\pi)}{\pi}$.
\end{cor}
\noindent
\emph{Proof.}
Lemma~\ref{lem:bound} together with $\varrho_\lambda(x)=J(x)$
and \eqref{Jx-final} gives the assertion.
\hspace*{\fill}$\square$%

\bigskip

The change of spectral dimension is important. If instead of \eqref{Fred}
the function $J$ was given by
$\tilde{J}(x)=x-\lambda x^2 \int_0^\infty dt\; \frac{\varrho_0(t)}{
(t+\mu^2)^2(t+\mu^2+x)}$, then for $\varrho_0(x)=x$
this function $\tilde{J}$ is bounded above. Hence, $\tilde{J}^{-1}$
needed in higher topological sectors could not exist globally on
$\mathbb{R}_+$, which would render the model inconsistent for any $\lambda>0$.
The dimension drop down to $D=4-2\frac{\arcsin(\lambda\pi)}{\pi}$
avoids this (triviality) problem.

\section{Perturbative expansion}

\label{Sec:perturbative}

In this section we study two different perturbative expansions
of an angle function which is behind the solution of $G(x,y)$.
In Sec.~\ref{Sec1} we directly expand
\eqref{winkeleq} order by order in $\lambda$, whereas in
Sec.~\ref{Sec2} we expand \eqref{Fred} and compare with the other
result via Corollary~\ref{cor:tau}. For a special
choice of $\mu^2$ which we determine, both expansions coincide order
by order in $\lambda$ (we played the game up to the $10^{\text{th}}$
order with a computer algebra system).

\subsection{Recalling earlier results}

\label{sec:earlier}

Equation \eqref{inteq} is a \textit{nonlinear} singular integral equation of Carleman type. The solution theory for
\textit{linear} integral equations is known (see e.g. \cite{Tricomi85})
and suggests the ansatz
\begin{align}\label{sol1}
 G(a,b)=\frac{\sin(\tau_b(a))}{\lambda\pi a}
 e^{\mathcal{H}_a^\Lambda[\tau_b(\bullet)]-\mathcal{H}_0^\Lambda[\tau_0(\bullet)]},
\end{align}
where
$ \mathcal{H}_a^\Lambda[f(\bullet)]:=\frac{1}{\pi}
 \lim_{\varepsilon\to0}\big(\int_0^{a-\varepsilon}
+\int_{a+\varepsilon}^{\Lambda^2}\big)
 dp\;\frac{f(p)}{p-a}$
denotes the finite Hilbert transform. Inserting \eqref{sol1} into
\eqref{inteq} gives with identities established 
in \cite{Panzer:2018tvy} the consistency relation
\begin{align}\label{winkeleq}
 p\lambda\pi \cot(\tau_a(p))=\mu_{bare}^2+a+p+\lambda \pi \mathcal{H}_p^\Lambda [\bullet] +
 \frac{1}{\pi} \int_0^{\Lambda^2} dt\, \tau_p(t).
\end{align}
Renormalisation by Taylor subtraction at $0$
suggests to choose the bare mass according to
\begin{align}\label{renormcond}
 \mu_{bare}^2=1-\lambda \Lambda^2-\frac{1}{\pi} \int_0^{\Lambda^2} dt\, \tau_0(t).
\end{align}
We will later see that another form of \eqref{renormcond} is for the
exact solution more efficient.

The key step in \cite{Grosse:2019jnv} to
solve \eqref{winkeleq} (actually in larger generality)
was to define a $\lambda$-deformation $\varrho_\lambda(x)$
of a spectral measure function $\varrho_0$.
This deformed measure then gives rise
to a function $J(x)$ which in four dimensions reads
\begin{align}\label{Jfunction}
 J(z):=z-\lambda z^2 \int_0^{\infty}dt\; \frac{{\varrho}_\lambda (t)}{
(t+\mu^2)^2 (t+\mu^2+z)}.
\end{align}
The system of functions $(\varrho_0,\varrho_\lambda,J)$ is
closed by the final condition  $\varrho_0(J(x))=\varrho_\lambda(x)$.

In general this is a complicated system of equations.
Here, the integral equation \eqref{inteq} encodes the spectral measure
$\varrho_0(x)=x$ so that $J(x)=\varrho_\lambda(x)$ and
\eqref{Jfunction} is reduced to \eqref{Fred}.
We now have the following corollary of \cite[Thm.\ 2.7]{Grosse:2019jnv}:
\begin{cor} \label{cor:tau}
Adjusting the bare mass to
\begin{align}
\label{mubare-2}
\mu^2_{bare}(\Lambda)= \mu^2\cdot \left(1-\lambda
\int_0^{J^{-1}(\Lambda^2)} dt\;
\frac{{\varrho}_\lambda (t)}{(t+\mu^2)^2 }\right)
-2\lambda \int_0^{J^{-1}(\Lambda)} dt\;
\frac{{\varrho}_\lambda (t)}{(t+\mu^2)},
\end{align}
then the consistency relation \eqref{winkeleq} is solved by
\begin{align}\label{Icot}
 \lambda \pi\varrho_0(p)\cot(\tau_a(p))
&=\lim_{\varepsilon\to0}\mathrm{Re}(a+I(p+\mathrm{i}\varepsilon)),
\\
\text{where}\qquad
I(z):=&-J(-\mu^2-J^{-1}(z)). \nonumber
\end{align}
\end{cor}
\noindent
Note that \eqref{mubare-2} fixes the renormalisation
different than \eqref{renormcond}. It is actually a family of
renormalisations which depend on a free parameter
$\mu^2(\lambda)$. Setting $G(0,0)=1$ does not
mean $\mu^2=1$, nevertheless both approaches coincide in the
limit $\Lambda^2\to \infty$. We will later identify this unique function
$\mu^2(\lambda)$ that gives \eqref{renormcond}.

\subsection{Direct expansion}\label{Sec1}

Expanding equation \eqref{winkeleq} with renormalisation \eqref{renormcond} and finite cut-off gives
\begin{align}\label{winkeleq3}
 p\lambda\pi \cot(\tau_a(p))=1+a+p+\lambda p\log\bigg(\frac{\Lambda^2-p}{p}\bigg)
 +\frac{1}{\pi}
 \int_0^{\Lambda^2} dt\left(\tau_p(t)-\tau_0(t)\right).
\end{align}
The first order is read out directly
\begin{align*}
 p\lambda\pi\cot(\tau_a(p))=1+a+p+\mathcal{O}(\lambda^1)\quad \Rightarrow\quad
 \tau_a(p)=\frac{p\lambda\pi}{1+a+p}+\mathcal{O}(\lambda^2),
\end{align*}
which gives after inserting back at the next order
\begin{align*}
 p\lambda\pi\cot(\tau_a(p))=&1+a+p+\lambda\bigg((1+p)\log(1+p)-p\log(p)\\
 &+p\log\left(\frac{\Lambda^2-p}{1+p+\Lambda^2}\right)
 +\log\left(\frac{1+\Lambda^2}{1+p+\Lambda^2}\right)\bigg)+\mathcal{O}(\lambda^2).
\end{align*}
The limit $\Lambda^2\to \infty$ gives finite results for
$\cot(\tau_a(p))$ as well as for $\tau_a(p)$ order by order, however
the limit has to be taken with caution. Integral and limit do
\textit{not} commute.  Namely, for and expansion
$\tau_a(p)=\sum_{n=1}^\infty \lambda^n\tau_a^{(n)}(p)$ we have
\begin{align*}
 \lim_{\Lambda^2\to\infty}\int_0^{\Lambda^2} dt\left(\tau_p^{(n)}(t)-\tau_0^{(n)}(t)\right)\neq
 \int_0^{\infty} dt\lim_{\Lambda^2\to\infty}\left(\tau_p^{(n)}(t)-\tau_0^{(n)}(t)\right), \qquad n>1.
\end{align*}
As an example we will look at the next order of both integrals. They give
\begin{align*}
  &\lim_{\Lambda^2\to\infty}\frac{1}{\pi}\int_0^{\Lambda^2} dt\left(\tau_p^{(2)}(t)-\tau_0^{(2)}(t)\right)\\
 &=(1+p)\log(1+p)^2+(1+2p)\mathrm{Li}_2(-p)-p\zeta_2,\\[10pt]
 &\frac{1}{\pi}\int_0^{\infty} dt\lim_{\Lambda^2\to\infty}\left(\tau_p^{(2)}(t)-\tau_0^{(2)}(t)\right)\\
 &=\int_0^{\infty} dt\,t\left(\frac{t\log(t)-(1+t)\log(1+t)}{(1+t+p)^2}-
 \frac{t\log(t)-(1+t)\log(1+t)}{(1+t)^2}\right)\\
 &=(1+p)\log(1+p)^2+(1+2p)\mathrm{Li}_2(-p)+2p\zeta_2,
\end{align*}
respectively, where $\mathrm{Li}_n(x)$ is the $n^{\text{th}}$
polylogarithm and $\zeta_n\equiv\zeta(n)$ is the Riemann zeta value at
integer $n$. The last term makes the difference.  Taking the ''wrong''
second result and plugging it back into \eqref{winkeleq3} would lead
to divergences at the next order. Consequently, we have to treat the
perturbative expansion of \eqref{winkeleq3} with a finite cut-off
$\Lambda^2$ at all orders, where each order has a finite limit.

The integration theory of the appearing integrals is completely understood
in form of iterated integrals \cite{Brown:2009qja}. They form a
shuffle algebra, which is symbolically implemented in the Maple
package \textsc{HyperInt} \cite{Panzer:2014caa}.

We computed the first 6 orders via \textsc{HyperInt} for finite
$\Lambda^2$. Sending $\Lambda^2\to\infty$ is well-defined
at any order as expected. The first orders read
explicitly
\begin{align}\nonumber
 \lim_{\Lambda^2\to\infty}p\lambda\pi\cot(\tau_a(p))&=1+a+p+\lambda\left((1+p)\log(1+p)-p\log(p)\right)\\
 &+\lambda^2\left(-p\zeta_2 +(1+p)\log(1+p)^2+(1+2p)\mathrm{Li}_2(-p)\right)\nonumber\\
 &+\lambda^3\big(\zeta_2 \log(1+p)- \frac{1+p}{2 p}\log(1+p)^2+(1+p)\log(1+p)^3\nonumber\\
 &\qquad +2p\zeta_3-2p\mathrm{Li}_3(-p)-(1+2p)\mathrm{Hlog}(p,[-1,0,-1])\nonumber\\
 &\qquad-2(2+3p)\mathrm{Hlog}(p,[0,-1,-1])\big)\label{perturbative}
 +\mathcal{O}(\lambda^4).
\end{align}
The hyperlogarithms Hlog are defined by the iterated integrals
\begin{align*}
 \mathrm{Hlog}(a,[k_1,...,k_n]):=&\int_0^a \frac{dx_1}{x_1-k_1}\int_0^{x_1}\frac{dx_2}{x_2-k_2}...
\int_0^{x_{n-1}}\frac{dx_n}{x_n-k_n},
\end{align*}
where the $k_i$ are called letters. An alternative notation is
$\mathrm{Hlog}(a,[k_1,...,k_n])=L_{k_1,...,k_n}(a)$. Important special cases are
$\mathrm{Hlog}(a,[\underbrace{-k,...,-k}_n])=
\frac{\log(1+\tfrac{a}{k})^n}{n!}$ for $k\in \mathbb{N}^\times $,
$\mathrm{Hlog}(a,[\underbrace{0,...,0}_n]):=\frac{\log(a)^n}{n!}$ and
$\mathrm{Hlog}(a,[\underbrace{0,...,0}_n,-1])=-\mathrm{Li}_{1+n}(-a)$.

The perturbative expansion shows that the branch point
at $p=-1$ plays an important role. Its boundary value is found to be
$
 \lim_{\substack{\Lambda^2\to\infty\\\varepsilon \searrow0}}
\cot(\tau_0(-1+\mathrm{i}\varepsilon))=
-\mathrm{i}+ \mathcal{O}(\lambda^7)$. It is natural to conjecture that it holds
at any order,
\begin{align}\label{conj}
 \lim_{\substack{\Lambda^2\to\infty\\ \varepsilon \searrow 0}}
\cot(\tau_0(-1+\mathrm{i}\varepsilon))=-\mathrm{i}.
\end{align}

The perturbative expansion with a finite cut-off $\Lambda^2$
is quite inefficient. The boundary value \eqref{conj} admits a more
efficient strategy.
We take the derivative of \eqref{winkeleq3} with respect to $p$:
\begin{align*}
 1+\lambda \log\Big(\frac{\Lambda^2{-}p}{p}\Big)
-\lambda\frac{\Lambda^2}{\Lambda^2{-}p}+\frac{1}{\pi}
 \int_0^{\Lambda^2}\!\!\! dt\;\frac{d\tau_p(t)}{dp}=
\lambda \pi \cot(\tau_a(p))+p \lambda\pi \frac{\partial}{\partial p}\cot(\tau_a(p)).
\end{align*}
Multiplying this equation by $p$ and subtracting it from \eqref{winkeleq3} again leads to
\begin{align}\label{AblFormel}
 -p^2\lambda\pi \frac{\partial}{\partial p}\cot(\tau_a(p))=
 1+a+\lambda\frac{p\Lambda^2}{\Lambda^2{-}p}
 +\frac{1}{\pi }\int_0^{\Lambda^2} \!\!\! dt
\left(\tau_p(t)-\tau_0(t)-p\frac{d\tau_p(t)}{dp}\right),
\end{align}
where the limit $\Lambda^2\to\infty$ is now safe from the
beginning and commutes with the integral.
We divide \eqref{AblFormel} by $-p^2$ and integrate it for
all orders higher than $\lambda^1$ over $p$ from
$-1$ (here  \eqref{conj} is assumed) up to some $q$ to get
$\lim_{\Lambda^2\to\infty} \lambda\pi \cot(\tau_a(q))$ on the lhs. On the rhs
the order of integrals $\int_{-1}^q dp\int_0^{\infty} dt$
can be exchanged. The integral over $p$ is
\begin{align}\label{intangl}
 \int_{-1}^q dp \frac{1}{p^2}\left(\tau_p(t)-\tau_0(t)
-p\frac{d\tau_p(t)}{dp}\right),
\end{align}
assuming H\"older continuity of $\tau_p(t)$ so that the integral splits after
taking principal values. The last term is
computed for small $\epsilon$ and all $\mathcal{O}(\lambda^{>1})$-contributions  via integration by parts
\begin{align}\nonumber
 \int_{[-1,q]\backslash [-\epsilon,\epsilon]}dp\frac{\frac{d\tau_p(t)}{dp}}{p}=&
 \frac{\tau_p(t)}{p}\bigg\vert_{p=\epsilon}^q+\frac{\tau_p(t)}{p}\bigg\vert_{p=-1}^{-\epsilon}
 +\int_{[-1,q]\backslash [-\epsilon,\epsilon]}dp\frac{\tau_p(t)}{p^2}\\
 =&\frac{\tau_q(t)}{q}+\tau_{-1}(t)+\int_{[-1,q]\backslash [-\epsilon,\epsilon]}dp\frac{\tau_p(t)}{p^2}
 -\frac{\tau_{-\epsilon}(t)+\tau_{\epsilon}(t)}{\epsilon}.\label{sitestep}
\end{align}
The first term in \eqref{intangl} cancels. The second term in
\eqref{intangl} integrates to a boundary term
$+2\frac{\tau_0(t)}{\epsilon}$, which is also canceled by the last term
of \eqref{sitestep}.
Multiplying by $q$ and including the special
$\mathcal{O}(\lambda)$-contribution we arrive in the limit
$\Lambda^2\to \infty$ where \eqref{conj} is (conjecturally) available at
\begin{align}\label{winkeleq4}
 q\lambda\pi \cot(\tau_a(q))=1+a+q-\lambda q \log(q)
 +\frac{1}{\pi}
 \int_0^{\infty} dt\left(\tau_q(t)-(1+q)\tau_0(t)+q \tau_{-1}(t)\right).
\end{align}
This equation is much more appropriate for the perturbation theory
because the number of terms is reduced tremendously order by
order. Obviously, the first six order coincide with the earlier but
much harder perturbative expansion of \eqref{winkeleq3}.

Using \eqref{winkeleq4} the perturbative expansion is increased up to
$\lambda^9$ with \textsc{HyperInt}.  As consistency check of
assumption \eqref{conj} we inserted
the next orders
$\tau_a^{(n)}(p)$ into \eqref{sol1} to get the expansion
$G(a,b)=\sum_{n=0}^\infty \lambda^n G^{(n)}(a,b)$. This confirmed
the symmetry $G^{(n)}(a,b)=G^{(n)}(b,a)$ which would easily be lost by
wrong assumptions. We are thus convinced to have
the correct expressions for $\tau_a^{(n)}(p)$ for $6<n<10$.

\subsection{Expansion of the Fredhom equation}
\label{Sec2}

To access the angle function $\tau_a(p)$ through Corollary~\ref{cor:tau}
we first have to determine the expansion of the
deformed measure ${\varrho}_\lambda(x)=J(x)$ through the Fredholm
equation \eqref{Fred}. The constant $\mu^2(\lambda)$ is not yet fixed and needs
a further expansion
\begin{align*}
 \mu^2=\sum_{n=0}^\infty \lambda^n\mu^2_n.
\end{align*}
First orders of the deformed measure are given iteratively through
\eqref{Fred}
\begin{align*}
 {\varrho}_\lambda(x)=&x-\lambda ((x+\mu^2_0)\Hlog(t,[-\mu^2_0])-x)\\
 &-\frac{\lambda^2}{\mu^2_0}(-\mu^2_0x\Hlog(x,[0,-\mu^2_0])+\mu^2_0(\mu^2_1+\mu^2_0+x)\Hlog(x,[-\mu^2_0])-x(\mu^2_1+\mu^2_0))\\
 &+\mathcal{O}(\lambda^3).
\end{align*}
Recall that the inverse of $J(x)={\varrho}_\lambda(x)=p$ exists for
all $p\in\mathbb{R}_+$ in case $\lambda< \left(\int_0^\infty
\frac{dt \,{\varrho}_\lambda (t)}{(t+\mu^2)^2 }\right)^{-1}$.
If $\varrho_\lambda(x)$ had the same asymptotics as $\varrho_0(x)=x$
then $J^{-1}$ could not be defined globally for $\lambda>0$.
We proved in sec. \ref{sec:spec}
that the asymptotics of $\varrho_\lambda(x)$ is
altered in such a way that $J^{-1}$ is defined.
Anyway, in each order of perturbative expansion the inverse $J^{-1}$
is globally defined on $\mathbb{R}_+$.
At this point it suffices to assume that $J^{-1}(p)$ is a formal power
series in $\lambda$, which is achieved by \eqref{Jfunction}
\begin{align*}
 J^{-1}(p)=p+\lambda (J^{-1}(p))^2\int_0^\infty\frac{dt\,{\varrho}_\lambda(t)}{(t+\mu^2)^2(t+\mu^2+J^{-1}(p))}.
\end{align*}
Expanding ${\varrho}_\lambda(t)$ and $\mu^2$, the first orders are
\begin{align*}
 J^{-1}(p)=&p-\lambda(p-(\mu^2_0+p)\Hlog(p,[-\mu^2_0]))\\
 &-\frac{\lambda^2}{\mu^2_0}(p\mu^2_0\Hlog(p,[0,-\mu^2_0])-2\mu^2_0(p+\mu^2_0)\Hlog(p,[-\mu^2_0,-\mu^2_0])\\
 &\qquad -\mu^2_0(\mu^2_1+\mu^2_0)\Hlog(p,[-\mu^2_0])+p(\mu^2_1+\mu^2_0))+\mathcal{O}(\lambda^3).
\end{align*}
The last step is to determine $\lim_{\varepsilon\to0}\mathrm{Re}I(p+i\varepsilon)=p\lambda\pi\cot(\tau_0(p))$ for $\Lambda^2\to \infty$
via
\begin{align*}
 I(z)=\mu^2+J^{-1}(z)+\lambda(\mu^2+J^{-1}(z))^2\int_0^\infty \frac{dt\,{\varrho}_\lambda(t)}{(t+\mu^2)^2(t-J^{-1}(z))},
\end{align*}
as a formal series.
The first few orders are
\begin{align*}
& \lim_{\varepsilon\to 0} I(p+i\varepsilon)
\\
&=\mu_0^2+p+\lambda(\mathrm{i}\pi p+\mu^2_0+\mu^2_1+(\mu^2_0+p)\Hlog(p,[-\mu^2_0])+p\log(\mu^2_0)-p\log(p)))\\
 &+\lambda^2\big(\mu_0^2(1-\zeta_2)+\mu_1^2+\mu_2^2-p\zeta_2
 +(\mu_0^2+\mu_1^2)\Hlog(p,[-\mu^2_0])\\
 &\quad +2(\mu_0^2+p)\Hlog(p,[-\mu^2_0,-\mu^2_0])-(\mu^2_0+2p)
 \Hlog(p,[0,-\mu^2_0])\big)+\mathcal{O}(\lambda^3).
\end{align*}
Comparing it with \eqref{perturbative} through equation \eqref{Icot}
fixes every $\mu^2_i$ uniquely and confirms
\begin{align*}
 \lim_{\varepsilon\to 0}I(p+\mathrm{i}\varepsilon)=\lambda\pi p\cot(\tau_0(p))+\mathrm{i}\lambda\pi p.
\end{align*}
Furthermore,
the first 10 orders are identical with the
expansion of \eqref{winkeleq4}, provided that the
$\mu^2_i$'s are fixed to
\begin{align}\label{mu}
 \mu^2=&1-\lambda+\frac{1}{6}(\pi\lambda)^2-\lambda \frac{1}{3}(\pi\lambda)^2
 +\frac{3}{40}(\pi\lambda)^4-\lambda\frac{8}{45}(\pi\lambda)^4
 +\frac{5}{112}(\pi\lambda)^6-\lambda\frac{4}{35}(\pi\lambda)^6\nonumber \\*
 &+\frac{35}{1152}(\pi\lambda)^8
 -\lambda\frac{128}{1575}(\pi\lambda)^8+\frac{63}{2816}(\pi\lambda)^{10}+\mathcal{O}(\lambda^{11}).
\end{align}
The conjectured behavior of $\cot(\tau_0(p))$ at $p=-1+\mathrm{i}\varepsilon$ in the previous subsection
\eqref{conj} is now equivalent to
\begin{align*}
 \lim_{\varepsilon\to0} I(-1+\mathrm{i}\varepsilon)=0\quad \Rightarrow\quad J^{-1}(-1)=-\mu^2.
\end{align*}
We find that
the expansion \eqref{mu} of $\mu^2$ obeys an unexpected boundary condition
\begin{align}\label{condmu}
 \int_0^\infty\frac{dt\,{\varrho}_\lambda(t)}{(\mu^2+t)^3}=\frac{1}{2}+\mathcal{O}(\lambda^{10}).
\end{align}

For further study we pass as in sec.~\ref{sec:DGL} to the rescaled measure
$\phi(x)=
\mu^2 \tilde{\varrho}_\lambda(\mu^2 x):=\frac{\varrho_\lambda(\mu^2 x)}{
\mu^2 x(1+ x)}$.
The pattern of coefficients of the $\mu^2$-expansion in \eqref{mu}
suggests to distinguish between even an odd powers in
$\lambda$. The even powers $\lambda^{2n}$ are given by the formula
\begin{align*}
 \frac{(2n-1)!!}{(2n)!! (2n+1)}=\frac{(2n)!}{4^nn!^2(2n+1)},
\end{align*}
and the odd powers $\lambda^{2n+1}$ by
\begin{align*}
 2\frac{(2n)!!}{(2n+1)!!(2n+2)}=2\frac{4^nn!^2}{(2n+2)!}.
\end{align*}
Both series are convergent for $|\lambda|<\frac{1}{\pi}$ with the result (up to order $\lambda^{10}$)
\begin{align*}
 \mu^2=\frac{\arcsin(\lambda\pi)}{\lambda\pi}-\lambda \left(\frac{\arcsin(\lambda\pi)}{\lambda\pi}\right)^2.
\end{align*}
This result suggests that $\frac{\arcsin(\lambda\pi)}{\pi}$ is a
better expansion parameter than $\lambda$ itself. The factors
$\pi^{2n}$ are produced by $\zeta_{2n}$ in the  iterated integrals.
We thus reorganise the perturbative solution of \eqref{feq}
into a series in $\frac{\arcsin(\lambda\pi)}{\pi}$.
The power of $\frac{\arcsin(\lambda\pi)}{\lambda\pi}$
depends on the
number of letters of the hyperpolylogarithm, which alternate between
 $-1$ and $0$.
The expansion which holds up to order $\lambda^{10}$ is given by
\begin{align}
\phi(x)=&c_\lambda \frac{\arcsin(\lambda\pi)}{\lambda\pi(1+x)}\sum_{n=0}^\infty \mathrm{Hlog}(x,[\underbrace{0,-1,...,0,-1}_{n}])
 \left(\frac{\arcsin(\lambda\pi)}{\pi}\right)^{2n}\\
 &-\lambda c_\lambda\frac{\arcsin(\lambda\pi)^2}{x(\lambda\pi)^2}\sum_{n=0}^\infty \mathrm{Hlog}(x,[-1,\underbrace{0,-1,...,0,-1}_{n}])
 \left(\frac{\arcsin(\lambda\pi)}{\pi}\right)^{2n},\nonumber
\end{align}
where the underbrace with $n$ means that we have $n$ times the letters $0$ and $-1$ in an alternating way.

In the limit $x\to0$ only the terms with $n=0$ in both sums survive,
\begin{align*}
1\equiv \phi(0)&=c_\lambda\frac{\arcsin(\lambda\pi)}{\lambda\pi}
\lim_{x\to0}\frac{\mathrm{Hlog}(x,[\,])}{1+x}-
 \lambda c_\lambda\frac{\arcsin(\lambda\pi)^2}{(\lambda\pi)^2}\lim_{x\to0}\frac{\mathrm{Hlog}(x,[-1])}{x}\\*
 &= \frac{c_\lambda}{\lambda}
\frac{\arcsin(\lambda\pi)}{\pi}
\Big(1-\frac{\arcsin(\lambda\pi)}{\lambda\pi}\Big).
\end{align*}
This value was found in sec.~\ref{sec:DGL} by another method.
We also remark that $c_\lambda=\frac{1}{\mu^2}$ for the special renormalisation.

Next define the functions
\begin{align*}
 f(x)&:=\sum_{n=0}^\infty \mathrm{Hlog}(x,[\underbrace{0,-1,...,0,-1}_{n}])
 \,\alpha_\lambda^{2n}\\
 g(x)&:=\sum_{n=0}^\infty\mathrm{Hlog}(x,[-1,\underbrace{0,-1,...,0,-1}_{n}])
 \,\alpha_\lambda^{2n},
\end{align*}
where $\alpha_\lambda=\frac{\arcsin(\lambda\pi)}{\pi}$.
Both together obey the differential equations
\begin{align*}
 f'(x)=\frac{\alpha_\lambda^2}{x}g(x)\qquad g'(x)=\frac{1}{1+x}f(x),
\end{align*}
or equivalently
\begin{align*}
 f''(x)+\frac{f'(x)}{x}-\alpha_\lambda^2 \frac{f(x)}{(1+x)x}=0,\quad
g''(x)+\frac{g'(x)}{1+x}-\alpha_\lambda^2 \frac{g(x)}{(1+x)x}=0,
\end{align*}
with the boundary conditions $f(0)=1$, $f'(0)=\alpha_\lambda^2$,
$g(0)=0$ and $g'(0)=1$.
The solution is given by hypergeometric functions $_2F_1$
\begin{align*}
 f(x)={}_2F_1\Big(\genfrac{}{}{0pt}{}{\alpha_\lambda,\;{-}\alpha_\lambda}{1}
\Big| -x\Big) \qquad
g(x)=\frac{x}{\alpha_\lambda^2}f'(x)=x{}_2F_1
\Big(\genfrac{}{}{0pt}{}{1{+}\alpha_\lambda,\;1{-}\alpha_\lambda}{2}
\Big| -x\Big).
\end{align*}
In summary. the solution of equation \eqref{feq} is conjectured to be
\begin{align}
\phi(x)&=\frac{\alpha_\lambda c_\lambda}{\lambda(1+x)}
{}_2F_1\Big(\genfrac{}{}{0pt}{}{\alpha_\lambda,\;{-}\alpha_\lambda}{1}
\Big| -x\Big)
 -\frac{\alpha_\lambda^2c}{\lambda}
{}_2F_1
\Big(\genfrac{}{}{0pt}{}{1{+}\alpha_\lambda,\;1{-}\alpha_\lambda}{2}
\Big| -x\Big)
\nonumber
\\
&={}_2F_1
\Big(\genfrac{}{}{0pt}{}{1{+}\alpha_\lambda,\;2{-}\alpha_\lambda}{2}
\Big| -x\Big)
\end{align}
or equivalently for \eqref{Fred}
\begin{align}
 J(x)=\varrho_\lambda(x)=&\frac{x}{\mu^2}\bigg(1+\frac{x}{\mu^2}\bigg)f\bigg(\frac{x}{\mu^2}\bigg)
=x  \,{}_2F_1
\Big(\genfrac{}{}{0pt}{}{\alpha_\lambda,\;1{-}\alpha_\lambda}{2}
\Big| -\frac{x}{\mu^2}\Big),
\label{Jx}
\end{align}
where we have used
the Gauss recursion formula \cite[$\S$ 9.137.7]{gradshteyn2007}
for hypergeometric functions.
Finally, we note that
\[
\int_0^\infty \frac{dt\; \varrho_\lambda(t)}{(t+\mu^2)^3}
=\lim_{x\to 0} \frac{x-\varrho_\lambda(x)}{\lambda x^2}
= \frac{\alpha_\lambda(1-\alpha_\lambda)}{2\lambda\mu^2}=\frac{1}{2c_\lambda\mu^2}\;.
\]
Thus choosing $\mu^2=\frac{\alpha_\lambda(1-\alpha_\lambda)}{\lambda}$
we confirm (\ref{condmu}) exactly.

\section{Stieltjes transform of the measure function}
\label{Sec:proof}

We find it interesting to directly check that the hypergeometric function
$\tilde{\varrho}_\lambda(x)=\frac{1}{\mu^2}\phi(\frac{x}{\mu^2})$,
see (\ref{phi-sol}), solves the integral equation (\ref{intc}).
The hypergeometric function can be expressed through the more general
Meijer-G function. A Meijer-G function is defined by
\begin{align}\label{AppDef}
 G^{m,n}_{p,q}\Big(z\Big|
\genfrac{}{}{0pt}{}{a_1,...,a_p}{b_1,...,b_q}\Big)=\frac{1}{2\pi \I}\int_L
\frac{\prod_{j=1}^m \Gamma(b_j-s)\prod_{j=1}^n \Gamma(1-a_j+s)}
{\prod_{j=m+1}^q \Gamma(1-b_j+s)\prod_{j=n+1}^p \Gamma(a_j-s)}z^sds,
\end{align}
with $m,n,p,q\in\N$, with $m\leq q$ and $n\leq p$, and
poles of
$\Gamma(b_j-s)$ different from poles of $\Gamma(1-a_j+s)$.
The infinite contour $L$ separates between the poles of
$\Gamma(b_j-s)$ and $\Gamma(1-a_j+s)$, and its behavior to
infinity depends on $m,n,p,q$ (see \cite[\S 9.3]{gradshteyn2007}).

The Meijer-G function has by definition the property
\begin{align}\label{AppInv}
 G^{m,n}_{p,q}\Big(z\Big|
\genfrac{}{}{0pt}{}{a_1,...,a_p}{b_1,...,b_q}\Big)=\frac{1}{z}G^{n,m}_{q,p}\Big(\frac{1}{z}\Big|
\genfrac{}{}{0pt}{}{-b_1,...,-b_q}{-a_1,...,-a_p}\Big).
\end{align}
It obeys the convolution formula \cite[$\S$ 7.811.1]{gradshteyn2007}
\begin{align}\nonumber
 &\int_0^\infty dx
 G^{m,n}_{p,q}\Big(\alpha x\Big|
\genfrac{}{}{0pt}{}{a_1,...,a_p}{b_1,...,b_q}\Big)
G^{m',n'}_{p',q'}\Big(\beta x\Big|
\genfrac{}{}{0pt}{}{a'_1,...,a'_{p'}}{b'_1,...,b'_{q'}}\Big)\\
&\qquad =\frac{1}{\alpha}G^{n+m' ,m+n'}_{q+p',p+q'}\Big(\frac{\beta}{\alpha}\Big|
\genfrac{}{}{0pt}{}{-b_1,..,-b_m,a'_1,..,a'_{p'},-b_{m+1},..,-b_q}
{-a_1,..,-a_n,b'_1,..,b'_{q'},-a_{n+1},..,-a_p}\Big),\label{AppConv}
\end{align}
which is the source of numerous impressive integrals over $\mathbb{R}_+$
of products of special functions.
If no two $b_j$ differ by an integer, either $p<q$ or $p=q$ with $|z|<1$,
then a Meijer-G function can
be expressed by hypergeometric functions
\begin{align}\label{AppHyp}
G^{m,n}_{p,q}\Big(z\Big|
\genfrac{}{}{0pt}{}{a_1,\dots,a_p}{b_1,\dots,b_q}\Big)
&=\sum_{k=1}^m\frac{\prod_{j=1}^{'m}\Gamma(b_j-b_k)\prod_{j=1}^n\Gamma(1+b_k-a_j)}
{\prod_{j=m+1}^q\Gamma(1+b_k-b_j)\prod_{j=n+1}^p\Gamma(a_j-b_k)}z^{b_k}
\\
&\times \,_pF_{q-1}
\Big(\genfrac{}{}{0pt}{}{1+b_k-a_1,\dots,1+b_k-a_p}{
1+b_k-b_1,..,\star,..,1+b_k-b_q}\Big|(-1)^{p-n-m}z\Big)\nonumber,
\end{align}
where primed sum and the $\star$ means that the term with $j=k$ is omitted.

We need another identity which is derived directly from the definition
\begin{align}\nonumber
 G^{3,2}_{3,3}\Big( z\Big|
\genfrac{}{}{0pt}{}{0,0,1}{b_1,b_2,0}\Big)=&\frac{1}{2\pi \I}\int_L
\frac{\Gamma(b_1-s)\Gamma(b_2-s)\Gamma(-s)\Gamma(1+s)^2}
{\Gamma(1-s)}z^sds
\\\nonumber
=&-\frac{1}{2\pi \I}\int_L
 \Gamma(b_1-s)\Gamma(b_2-s)\Gamma(s)\Gamma(1+s)
z^sds
\\\nonumber
=&\Gamma(b_1)\Gamma(b_2)-\frac{1}{2\pi \I}\int_{L'}
 \Gamma(b_1-s)\Gamma(b_2-s)\Gamma(s)\Gamma(1+s)
z^sds\\
=&\Gamma(b_1)\Gamma(b_2)-G^{2,2}_{2,2}\Big( z\Big|
\genfrac{}{}{0pt}{}{0,1}{b_1,b_2}\Big)\label{AppSpec},
\end{align}
where the contour is changed $L\to L'$ such that it is moved
through $s=0$ and picked up the residue.
The contour $L'$ fulfills the definition \eqref{AppDef} for $G^{2,2}_{2,2}\Big( z\Big|
\genfrac{}{}{0pt}{}{0,1}{b_1,b_2}\Big)$.

From \eqref{AppHyp} one can establish
\begin{align*}
\tilde{\varrho}_\lambda(t)&=
\frac{1}{\mu^2}
\frac{1}{\Gamma(2-\alpha_\lambda)\Gamma(1+\alpha_\lambda)}
G^{1,2}_{2,2}\Big(\frac{t}{\mu^2}\Big|
\genfrac{}{}{0pt}{}{\alpha_\lambda-1,-\alpha_\lambda}{0,-1}\Big)\;,
\end{align*}
and $\frac{1}{x+t+\mu^2}= \frac{1}{x+\mu^2}
{}_1F_0\big(\genfrac{}{}{0pt}{}{1}{-}\big|
{-}\frac{t}{x+\mu^2}\big)=
\frac{1}{x+\mu^2}
G^{1,1}_{1,1}\big(\frac{t}{x+\mu^2}\big|
\genfrac{}{}{0pt}{}{0}{0}\big)$.
The convolution theorem \eqref{AppConv} of
Meijer-G functions thus allows to evaluate the integral
\begin{align}
&\lambda
\int_0^\infty \frac{dt\;\tilde{\varrho}_\lambda(t)}{x+t+\mu^2}
\nonumber
\\
&\stackrel{\text{\eqref{AppConv}}}{=} \frac{\lambda}{\mu^2\Gamma(2-\alpha_\lambda)\Gamma(1+\alpha_\lambda)}
G^{2,3}_{3,3}\Big(\frac{x+\mu^2}{\mu^2}\Big|
\genfrac{}{}{0pt}{}{\alpha_\lambda-1,-\alpha_\lambda,0}{0,0,-1}\Big)
\nonumber
\\
&\stackrel{\text{\eqref{AppInv}}}{=}
\frac{\lambda}{(x+\mu^2)\Gamma(2-\alpha_\lambda)\Gamma(1+\alpha_\lambda)}
G^{3,2}_{3,3}\Big(\frac{\mu^2}{x+\mu^2}\Big|
\genfrac{}{}{0pt}{}{0,0,1}{1-\alpha_\lambda,\alpha_\lambda,0}\Big)
\nonumber
\\
&\stackrel{\text{\eqref{AppSpec}}}{=}
\frac{\lambda}{(x+\mu^2)\Gamma(2-\alpha_\lambda)\Gamma(1+\alpha_\lambda)}\left(\Gamma(1-\alpha_\lambda)
\Gamma(\alpha_\lambda)-
G^{2,2}_{2,2}\Big(\frac{\mu^2}{x+\mu^2}\Big|
\genfrac{}{}{0pt}{}{0,1}{1-\alpha_\lambda,\alpha_\lambda}\Big)\right)
\nonumber
\\
&\stackrel{\text{\eqref{AppHyp}}}{=}
\frac{\lambda}{(x+\mu^2)}
\Big\{\frac{1}{\alpha_\lambda(1-\alpha_\lambda)}
\nonumber
\\
&-
\frac{\Gamma(2\alpha_\lambda-1)
\Gamma(1-\alpha_\lambda)}{\Gamma(1+\alpha_\lambda)}
\Big(\frac{\mu^2}{x+\mu^2}\Big)^{1-\alpha_\lambda}
\;{}_2F_1
\Big(\genfrac{}{}{0pt}{}{2{-}\alpha_\lambda,\;1{-}\alpha_\lambda}{2{-}2\alpha_\lambda}
\Big| \frac{\mu^2}{x+\mu^2}\Big)
\nonumber
\\
& -\frac{\Gamma(1-2\alpha_\lambda)
\Gamma(\alpha_\lambda) }{\Gamma(2-\alpha_\lambda)}
\Big(\frac{\mu^2}{x+\mu^2}\Big)^{\alpha_\lambda}
\;{}_2F_1
\Big(\genfrac{}{}{0pt}{}{1{+}\alpha_\lambda,\;\alpha_\lambda}{2\alpha_\lambda}
\Big| \frac{\mu^2}{x+\mu^2}\Big)
\Big\}
\nonumber
\\
&=
\frac{1}{(x+\mu^2)}
\frac{\lambda}{\alpha_\lambda(1-\alpha_\lambda)}
- \frac{\lambda\pi}{\sin (\alpha_\lambda \pi)} \tilde{\varrho}_\lambda(x)\;.
\end{align}
We have used the expansion of a Meijer-G function into hypergeometric
functions and applied in the last step  \cite[\S 9.132.1]{gradshteyn2007}.
The result is precisely (\ref{intc})
provided that
$c_\lambda=\frac{\lambda}{\alpha_\lambda(1-\alpha_\lambda)}$ (see \eqref{phi-sol}) and
$\sin( \alpha_\lambda\pi)=\lambda\pi$ (see \eqref{sol-alpha}).

\section{Outlook}

With the identification of $J$ we have completed the solution of the
planar $2$-point function of the $\Phi^4$-QFT model on four-dimensional
Moyal space at the self-duality point. From the 2-point function one
directly builds all planar correlation
functions \cite{Grosse:2012uv, deJong}.
In our earlier work \cite{Grosse:2019nes} on
the much simpler cubic Kontsevich-like model we gave an algorithm to
compute also all non-planar correlation functions from the planar sector.
It remains to be seen whether a similar endeavour can succeed
for the $\Phi^4$-model, too.

We expect that non-planar functions are expressed in
terms of the inverse function $J^{-1}$. Inverses of hypergeometric functions
do not seem to be studied. There is now strong motivation to try it.
Of course one can approximate $J^{-1}$ perturbatively via the expansion
of $J$ into hyperlogarithms which we established. A non-perturbative
characterisation of $J^{-1}$ could provide useful identities between these
number-theoretic functions.

\section*{Acknowledgements}

This work was supported by the Erwin
Schr\"odinger Institute \mbox{(Vienna)} through a ``Research in Team''
grant and by the Deutsche Forschungsgemeinschaft via the Cluster of
Excellence\footnote{``Gef\"ordert durch die Deutsche
  Forschungsgemeinschaft (DFG) im Rahmen der Exzellenzstrategie des
  Bundes und der L\"ander EXC 2044 - 390685587, Mathematik M\"unster:
  Dynamik--Geometrie--Struktur"} ``Mathematics M\"unster'' and the RTG
2149.

%%%%%%%%%%%%%%%%%%%%%%%%%%%%%%%%%%%%%%%%%%%%%%%%%%%%%%%%%%%%%%%%%%%%%%

\appendix

\section[On the spectrum of the integral operator]{On the
spectrum of the integral operator
\\ (by Robert Seiringer)}

\label{app}

% The equation to be solved is
% $$
% \rho(t) = t - \lambda t^2 \int_0^\infty \frac{\rho(u)}{(u+m)^2(u+t+m)} du
% $$
% for some $m>0$. To make it look more symmetric, we write $\rho(u) = \psi(u) u(u+m)$ and rewrite the equation as
% $$
% \psi(t) = \frac 1{t+m} - \lambda \int_0^\infty \psi(u) \frac {u t}{(u+m)(u+t+m)(t+m)} du
% $$

Abstractly, the integral equation (\ref{intc}) is of the form
\[
\psi = f_\mu - \lambda A_\mu \psi,
\]
where $\psi(t)=\tilde{\varrho}_\lambda(t)$,
$f_\mu(t)= (t+\mu^2)^{-1}$ and $A_\mu$ is the operator with integral kernel
\begin{align}
A_\mu(t,u)=  \frac {u t}{(u+\mu^2)(u+t+\mu^2)(t+\mu^2)}.
\end{align}
Note that $A_\mu$ is symmetric and positive. The equation can thus be
solved for $\psi$ if $\lambda > \lambda_c = -\|A_\mu\|^{-1}$.

By scaling, the spectrum of $A_\mu$ is independent of $\mu$ for $\mu>0$.
We claim that
\begin{align}
\|A_\mu\| = \|A_0\| = \pi.
\end{align}
In particular, $\lambda_c = -1/\pi$.

Since $A_\mu$ has a positive kernel which is monotone in $\mu$, one
readily obtains $\|A_\mu\|\leq \|A_0\|$. On the other hand, $A_0$ is the
weak limit of $A_\mu$ as $\mu\to 0$, hence $\|A_0\|\leq \liminf_{\mu\to
  0}\|A_\mu\|$, which proves that $\|A_\mu\|=\|A_0\|$. Now $A_0(t,u)=
(u+t)^{-1}$. Introducing logarithmic coordinates, we have
\begin{align}
\int_0^\infty\int_0^\infty \frac{ \phi(u)^* \phi(t)}{u+t} du dt
&=
\int_{\mathbb{R}} \int_{\mathbb{R}} \frac{\phi^*(e^v) \phi(e^s)}{e^v + e^s}
e^{v+s} dv ds
\nonumber
\\
&=  \int_{\mathbb{R}} \int_{\mathbb{R}}
\frac{\phi^*(e^v)e^{v/2} \phi(e^s)e^{s/2}}{2 \cosh( \tfrac 12(v-s) )} dv ds
\end{align}
which can be diagonalised via Fourier transforms. Since
\[
\int_{\mathbb{R}} \frac 1 {2 \cosh(v/2) } dv = \pi,
\]
this shows that the spectrum of $A_0$ equals $[0,\pi]$, and
indeed $\|A_0\|=\|A_\mu\|=\pi$.

\addcontentsline{toc}{section}{References}
\bibliographystyle{halpha-abbrv}
\bibliography{./Bibo}

\end{document}